\documentclass[preprint, double-spaced]{revtex4}
\usepackage{graphicx}
\usepackage{amsmath,bbm}
\usepackage{amssymb,bm}

\begin{document}

\title{Particle Production at CBM in a Thermal Model Approach}
\author{A. Prakash}
\author{P.~K.~Srivastava\footnote{corresponding author: $prasu111@gmail.com$}}
\author{B.~K.~Singh}
%\author{C.~P.~Singh}

\affiliation{Department of Physics, Banaras Hindu University, 
Varanasi 221005, INDIA}

\begin{abstract}
\noindent
The Compressed Baryonic Matter (CBM) experiment planned at Facility for Antiproton and Ion Research (FAIR) will provide a major scientific effort for exploring the properties of strongly interacting matter in the high baryon density regime. One of the important goal behind such experiment is to precisely determine the equation of state (EOS) for the strongly interacting matter at extreme baryon density. In this paper, we have used a thermal model EOS incorporating excluded volume description for the hot and dense hadron gas (HG). We then predict different particle ratios and the total multiplicity of various hadrons in the CBM energy range i.e. from $10$ A GeV to $40$ A GeV lab energies, which corresponds to $4.43$ A GeV and $8.71$ A GeV center-of-mass energies. Our main emphasis is to estimate the strange particles enhancement as well as increase in the net baryon density in CBM experiment. We have also compared our results with the results obtained from various other theoretical approaches existing in the literature such as hadron string dynamics (HSD) model and ultra-relativistic quantum molecular dynamics (UrQMD) etc.
\\

 PACS numbers: 12.38.Mh, 12.38.Gc, 25.75.Nq, 24.10.Pa

\end{abstract}
\maketitle 
\section{ Introduction}
Relativistic heavy ion collisions aim at creating matter at extreme conditions of temperature ($T$) and/or baryon chemical potential ($\mu_{B}$). The theory of strong interactions i.e., Quantum Chromodynamics (QCD) predicts that at such conditions hadron gas (HG) will make a phase transition to a state governed by the partonic degrees of freedom called Quark-Gluon Plasma (QGP). In recent years, a number of experimental as well as theoretical attempts are being made to study the formation and detection of the properties and signals of QGP~\cite{cpsingh,cpsingh2}. We expect that different quark flavours are abundantly produced in the hot,dense region generated by the relativistic heavy ion collisions and particularly in the region at large $\mu_{B}$; one should get an enhancement of strange ($s$) and anti-strange ($\bar{s}$) quarks. Strangeness enhancement has been proposed as one of the early and important signals of QGP formation~\cite{rafelski,rafe,cp,vkt,cp1}. It has been argued that if a quark gluon plasma is formed from compressed nuclear matter as may happen in the nuclear fragmentation region and/or in the low energy ``stopping regime'' collisions, then the abundance of $s$ and $\bar{s}$ quark would be highly enhanced compared to that of light $u$ or $d$ quark~\cite{rafelski,rafe,biro,biro1}. This is possibly due to the Pauli exclusion principle which strongly suppresses the creation of light-quark pairs~\cite{margeti}. This asymmetry in the flavour composition generated by a baryon-rich QGP should result in a large production of $K^{+},~K^{-},~\Lambda,~\bar{\Lambda}$ etc~\cite{cp2}. This effect is even more evident in the case of the multistrange hyperons. A striking observation reported by NA49 collaboration is a pronounced and sharp maximum in the excitation function of $K^{+}/\pi^{+}$ ratio at 30 AGeV~\cite{alt1}. This sharp maximum which is also known as ``horn'', is not seen in p+p collisions. As $K^{+}$ is by far most abundant carrier of anti-strangeness at SPS energies, it also provides a good measure of the total strangeness produced in the collision. The ratio $K^{+}/\pi^{+}$ represents the strangeness to entropy ratio. A sharp maximum in this qunatity was predicted by the statistical model indicating the early stage as a consequence of the transistion to a deconfined state~\cite{cp2}. A similar maximum at the same beam energy is  also reported by the same collaboration for other strange particles like $\Lambda$'s and $\Xi^{-}$~\cite{alt2}. These observations confirm that this particular feature is not given by $K^{+}$ alone, but represent the total strangeness content of the final state~\cite{friese}. The measurement of the excitation function of strangeness production by NA49 collaboration have renewed a fresh stimulating discussion about the role of strangeness as a signature for the deconfinement phase transistion.

Similar to the strangeness enhancement, a large production of anti-baryons with respect to baryons are also proposed as the signal for the formation of deconfined QGP~\cite{i,j}. In heavy ion collisions, the system has a non-zero baryon number density due to nuclear stopping. At small and moderate center-of-mass energies as existing in the case of CBM experiment, the nuclear stopping is large in comparison with RHIC and/or LHC energies where nuclear transparency is found to dominate. Nuclear stopping leads to an asymmetry between the production of hadrons and antihadrons since the baryon number is conserved due to $U(1)$ global symmetry of the QCD Lagrangian and one would not expect an additional asymmetry in the produced particles other than the initial finite baryon number. However, there is a possibility that hadronization can also generate additional particle-antiparticle asymmetry. This asymmetry can be measured by the ratio of yields of antihadrons to hadrons~\cite{G,G1}. Ratios of yield of antiprotons to protons ($\bar{p}/p$) and that of antikaons to kaons ($K^{-}/K^{+}$) are the representatives of two such significant observables measuring the hadron-antihadron asymmetry in heavy ion collisions~\cite{G,G1}. The ratio $\bar{p}/p$ carries the information regarding baryons-antibaryons asymmetry and the ratio $K^{-}/K^{+}$ almost cancels the effect of strangeness production and indicates the asymmetry between charged mesons and their antiparticles generated in the hot, dense medium. 

 The experimental discovery of QGP would be a major breakthrough in our current understanding of the properties of nuclear matter. Therefore, several experimental programs were planned to explore the properties of strongly interacting matter and to search the possible existence of QGP e.g. relativistic heavy ion collider (RHIC) experiment at BNL~\cite{star,star1,phobos}, super-proton-synchrotron (SPS) at CERN~\cite{na,na57} etc. The Compressed baryonic matter (CBM) experiment at Facility for antiproton and ion research (FAIR) machine will provide the similar hot, dense situtaion in the laboratory to explore the enhancement of strange hadrons with respect to light hadrons and the enhancement of anti-baryons over baryons along with other signatures suitable for the detection of QGP production~\cite{senger}. However, after hadronization of the partonic plasma, we have a hot and dense HG. In this context, the search for a proper equation of state is of extreme importance because it can suitably describe the properties of hot and dense HG. A large number of thermal models was used in the recent past to deduce the multiplicities and ratios of particles emerging from the equilibrated HG at chemical freezeout and their agreements with the experimental data were found to be excellent~\cite{andro,cley,beca,bronio}. 

 Recently we have proposed a statistical thermal model for EOS of a hot, dense HG which incorporates the excluded volume effect due to finite hard-core size of the baryons~\cite{skt,pks} in a thermodynamically consistent manner. This EOS suitably describes the lattice results regarding thermodynamical and transport properties of HG phase at zero as well as at finite $\mu_{B}$~\cite{pks,pks1}. In this paper, our motivation is to precisely determine the multiplicity and the ratios of various hadrons in the CBM energy range. We mainly emphasize on the net baryon density created at freezeout, the production of strange particles and the asymmetry in the particle-antiparticle production etc. because these isuues are yet to be resolved in order to get information on the QCD phase transition. 

\noindent
\section{EOS for a Hadron gas}
Recently we have proposed a thermodynamically consistent excluded volume model for the hot and dense hadron gas (HG). In this model, the grand canonical partition function for the HG with full quantum statistics and after  suitably incorporating excluded volume correction is~\cite{skt,pks,pks1}:

\begin{equation}
\begin{split}
ln Z_i^{ex} = \frac{g_i}{6 \pi^2 T}\int_{V_i^0}^{V-\sum_{j} N_j V_j^0} dV
\\
\int_0^\infty \frac{k^4 dk}{\sqrt{k^2+m_i^2}} \frac{1}{[exp\left(\frac{E_i - \mu_i}{T}\right)+1]}
\end{split}
\end{equation}
where $g_i$ is the degeneracy factor of ith species of baryons,$E_{i}$ is the energy of the particle ($E_{i}=\sqrt{k^2+m_i^2}$), $V_i^0$ is the eigenvolume of one baryon of ith species and $\sum_{j}N_jV_j^0$ is the total occupied volume and $N_{j}$ represents total number of baryons of jth species.

Now we can write Eq.(1) as:

\begin{equation}
ln Z_i^{ex} = V(1-\sum_jn_j^{ex}V_j^0)I_{i}\lambda_{i},
\end{equation}
where $I_{i}$ represents the integral:
\begin{equation}
I_i=\frac{g_i}{6\pi^2 T}\int_0^\infty \frac{k^4 dk}{\sqrt{k^2+m_i^2}} \frac1{\left[exp(\frac{E_i}{T})+\lambda_i\right]},
\end{equation}
and $\lambda_i = exp(\frac{\mu_i}{T})$ is the fugacity of the particle, $n_j^{ex}$ is the number density of jth type of baryons after excluded volume correction and can be obtained from Eq.(2) as:
\begin{equation}
n_i^{ex} = \frac{\lambda_i}{V}\left(\frac{\partial{ln Z_i^{ex}}}{\partial{\lambda_i}}\right)_{T,V}
\end{equation}
This leads to a transcendental equation :
\begin{equation}
n_i^{ex} = (1-R)I_i\lambda_i-I_i\lambda_i^2\frac{\partial{R}}{\partial{\lambda_i}}+\lambda_i^2(1-R)I_i^{'}
\end{equation}
where $I_{i}^{'}$ is the partial derivative of $I_{i}$ with respect to $\lambda_{i}$ and $R=\sum_in_i^{ex}V_i^0$ is the fractional occupied volume depending on $n_{i}^{ex}$. We can write R in an operator equation as follows~\cite{skt}:
\begin{equation}
R=R_{1}+\hat{\Omega} R
\end{equation}
where $R_{1}=\frac{R^0}{1+R^0}$ with $R^0 = \sum n_i^0V_i^0 + \sum I_i^{'}V_i^0\lambda_i^2$; $n_i^0$ is the density of pointlike baryons of ith species and the operator $\hat{\Omega}$ has the form :
\begin{equation}
\hat{\Omega} = -\frac{1}{1+R^0}\sum_i n_i^0V_i^0\lambda_i\frac{\partial}{\partial{\lambda_i}}
\end{equation}
Using Neumann iteration method and retaining the series upto $\hat{\Omega}^2$ term, we get
\begin{equation}
R=R_{1}+\hat{\Omega}R_{1} +\hat{\Omega}^{2}R_{1}
\end{equation}
\noindent
Eq.(8) can be solved numerically. Finally, we get the total pressure~\cite{skt,pks} of the hadron gas as:
\begin{equation}
\it{p}_{HG}^{ex} = T(1-R)\sum_iI_i\lambda_i + \sum_i\it{P}_i^{meson}
\end{equation}

In Eq. (9), the first term represents the pressure due to all types of baryons where excluded volume correction is incorporated and the second term gives the total pressure due to all mesons in HG having a pointlike size. This makes it clear that we consider the hard-core repulsion arising between two baryons only. Essentially we consider that the mesons can interpenetrate each other but baryons can not owing to their hard-core size. In this calculation, we have taken an equal volume $V^{0}=\frac{4 \pi r^3}{3}$ for each type of baryon with a hard-core radius $r=0.8 fm$. We have taken all baryons and mesons and their resonances having masses upto $2 GeV/c^{2}$ in our calculation for the HG pressure. We have also imposed the condition of strangeness conservation by putting $\sum_{i}S_{i}(n_{i}^{s}-\bar{n}_{i}^{s})=0$, where $S_{i}$ is the strangeness quantum number of the ith hadron, and $n_{i}^{s}(\bar{n}_{i}^{s})$ is the strange (anti-strange) hadron density, respectively. Using this constraint equation, we get the value of strange chemical potential in terms of $\mu_{B}$. 

In order to relate the thermal parameters of hot,dense HG with the center-of-mass energy, we extract them by fitting the experimental particle ratios from lowest SIS energy to the highest RHIC energy by our model calculation. We then parametrize the variables $T$ and $\mu_{B}$ in terms of $\sqrt{s_{NN}}$ as given in Ref.~\cite{skt}.
\begin{figure}
\begin{center}
\noindent
\includegraphics[height=28em]{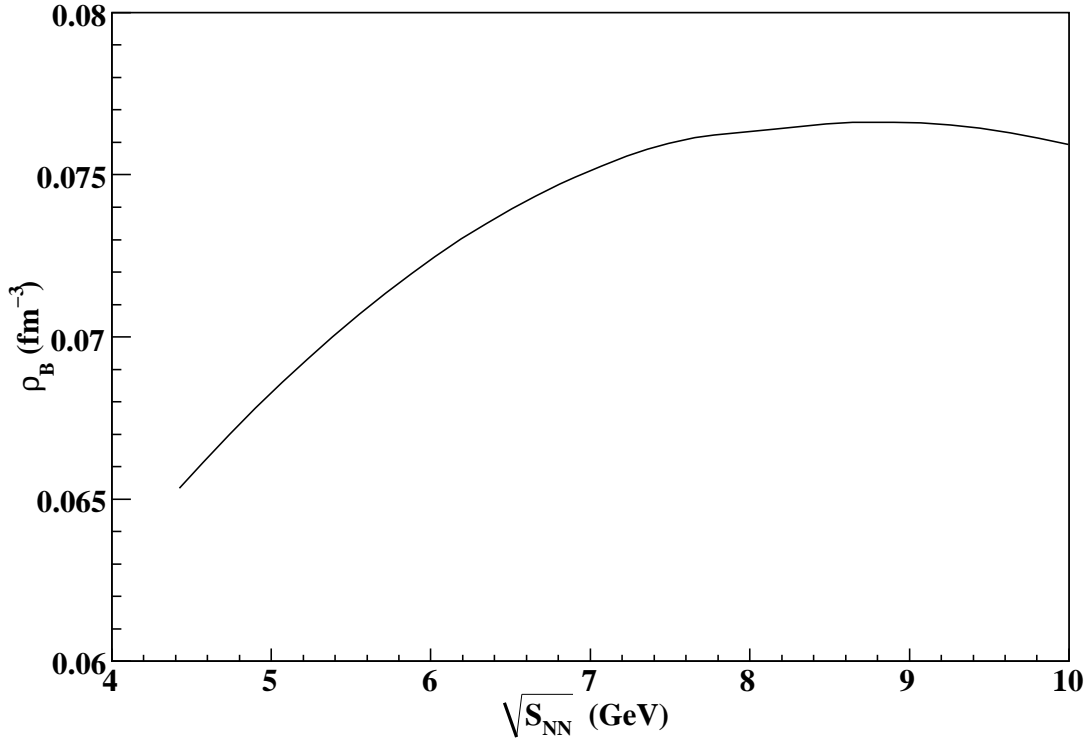}%use pdflatex for pdf
\caption[]{Variation of net-baryon density ($\rho_{B}$) at freezeout with respect to center of mass energy ($\sqrt{s_{NN}}$).}
\end{center}
\end{figure}
\begin{figure}
\begin{center}
\noindent
\includegraphics[height=28em]{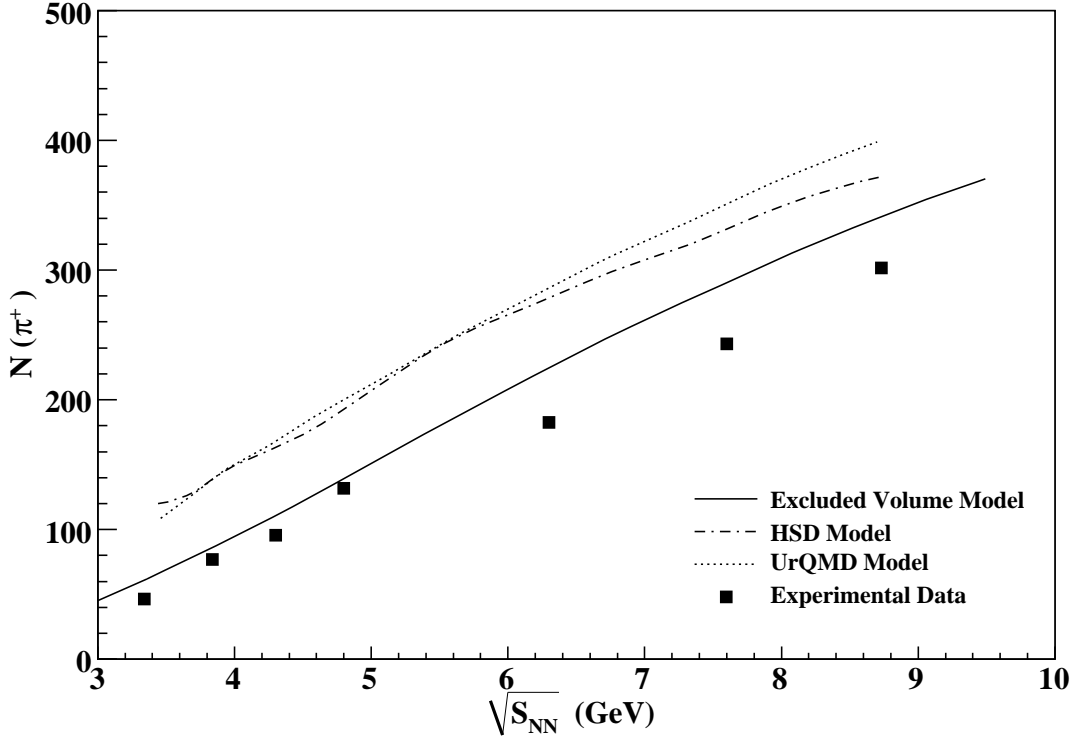}%use pdflatex for pdf
\caption[]{Variation of total multiplicity of produced $\pi^{+}$ with respect to $\sqrt{s_{NN}}$. Dash-dotted and dotted curve is the results obtained from HSD and UrQMD model, respectively~\cite{hsd}. Experimental data is taken from Ref.~\cite{eleven,alt1,alt2,blume1,afan,gaz}.}
\end{center}
\end{figure}

\section{Results and Discussions}
The extracted freezeout temperature in statistical thermal models of HG generally increases monotonically with the collision energy. However, the corresponding net baryon density exhibits a more complicated behaviour~\cite{randrup}. In the present excluded volume model, the net baryon density increases with $\sqrt{s_{NN}}$ (see Fig. 1), reaches  a maximum value near $\sqrt{s_{NN}}=8-9$ GeV and then decreases. The maximum freezeout net baryon density is approximately half of the normal nuclear density $\rho_{0}=0.15~fm^{-3}$. The maximum value of $\rho_{B}$ obtained in our model is lower than the value obtained in Ref.~\cite{randrup} where a HRG model is used without any excluded volume correction. Thus excluded volume effect shifts the net baryon density achieved at freezeout to a lower value. One important observation is that the CBM experiment can create a hadronic fireball system at freezeout having almost maximum achievable net baryon density by heavy-ion collisions. 

\begin{figure}
\begin{center}
\noindent
\includegraphics[height=28em]{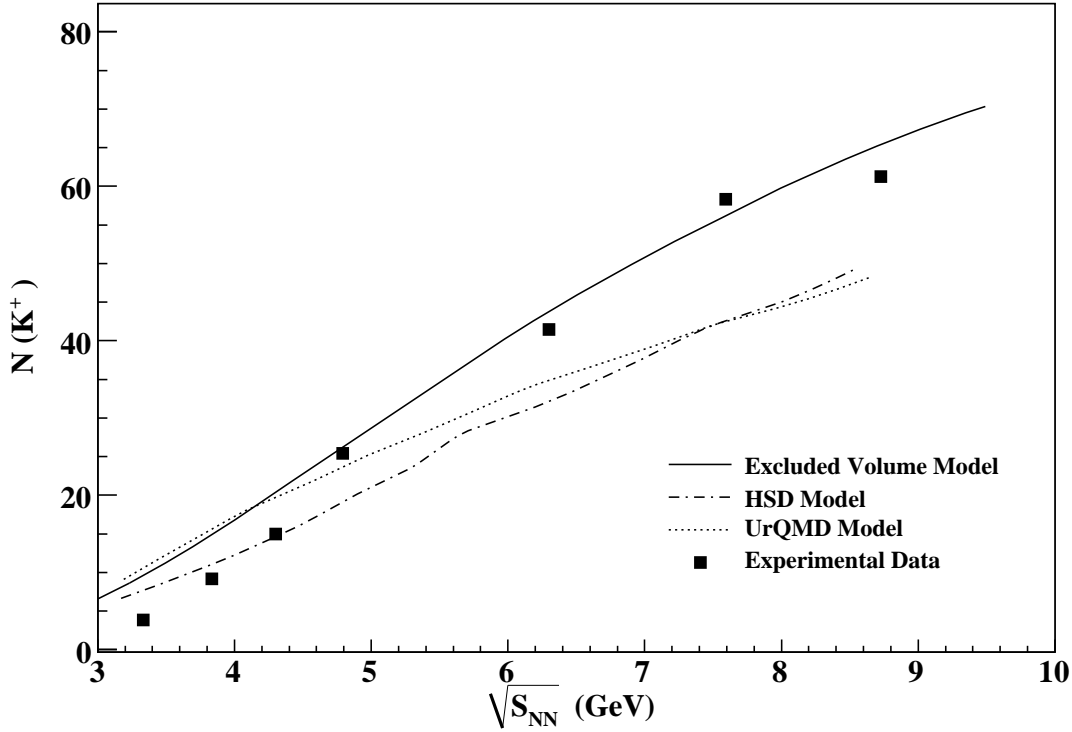}%use pdflatex for pdf
\caption[]{Variation of total multiplicity of produced $K^{+}$ with respect to $\sqrt{s_{NN}}$. Dash-dotted and dotted curve is the results obtained from HSD and UrQMD model, respectively~\cite{hsd}. Experimental data is taken from Ref.~\cite{eleven,alt1,alt2,blume1,afan,gaz}.}
\end{center}
\end{figure}

Fig. 2 presents the variation of total multiplicity of $\pi^{+}$ with respect to $\sqrt{s_{NN}}$. In the CBM energy range the fireball volume, at freezeout, extracted in the excluded volume model approach appears almost constant~\cite{skt1}. We have taken $5000$ $fm^{3}$ as the fireball volume in order to calculate the total multiplicity of hadrons. We compare our results with the experimental data obtained at AGS and SPS~\cite{eleven,alt1,alt2,blume1,afan,gaz} at low energies. We have also shown the total multiplicity of $\pi^{+}$ obtained from transport models like HSD and UrQMD. Both HSD~\cite{hsd} and UrQMD~\cite{qmd1,qmd2} models usually employ the concepts of string, quark, diquark, ($q,~\bar{q},~qq,~\bar{q}\bar{q}$) as well as the hadronic degrees of freedom. However, the numerical evaluations are quite different in HSD as compared to UrQMD. The UrQMD includes all baryonic resonances upto an invariant mass of $2$ GeV as well as mesonic resonances upto $1.9$ GeV~\cite{qmd1,qmd2}. However, HSD incorporates only the baryon octet and decuplet states and $N^{*}~(1440),~N^{*}~(1535)$ as well as their antiparticles together with the $0^{-}$ and $1^{-}$ meson octets. Higher baryonic resonances are discarded as the resonance structure (above $\Delta$ peak) is not clearly seen experimentally even in the photo-absorption by light nuclei~\cite{delta}. In contrast to the UrQMD at low energy baryon-baryon and meson-baryon collisions, HSD includes the direct (nonresonant) meson production. Our excluded volume model suitably describes the data upto $5$ GeV. However, all three models yield larger multiplicity for $\pi^{+}$ relative to the data above $5$ GeV. Although, the multiplicity of the produced $\pi^{+}$ in excluded volume model is closer to the data in comparison to other two models. In principle, the yield of each particle is mainly governed by the particle fugacity essentially determined from the chemical freeze-out parameters. It also depends on the size of the system (or volume $V$) in which some variations can occur.
\begin{figure}
\begin{center}
\noindent
\includegraphics[height=28em]{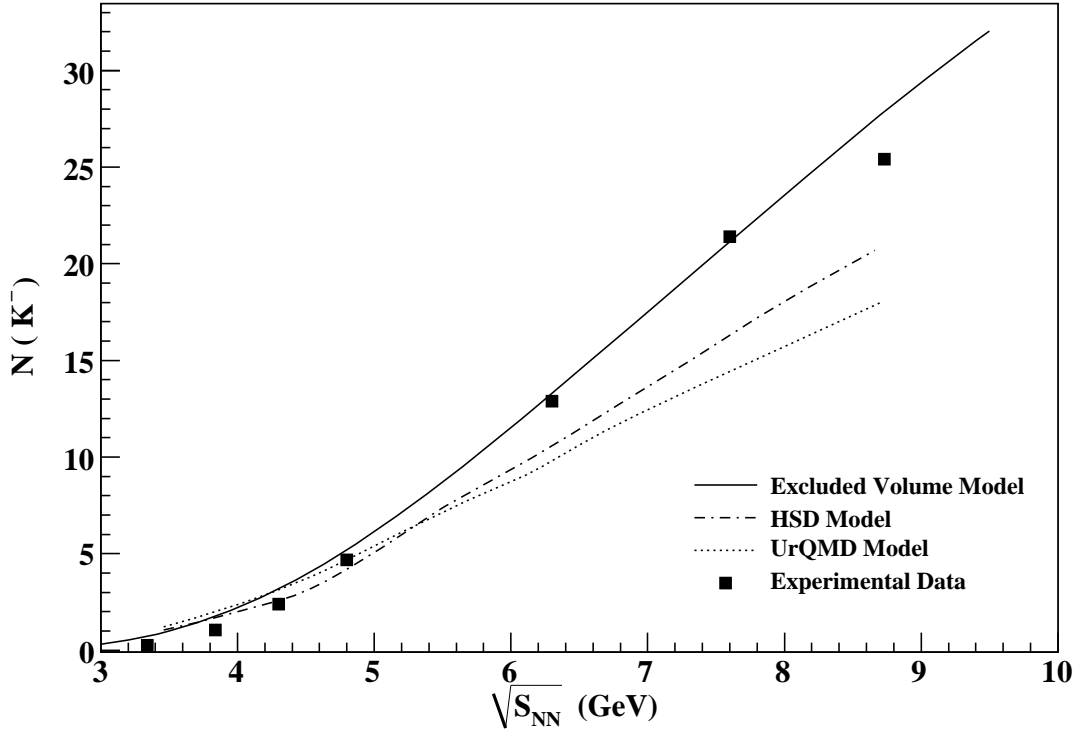}%use pdflatex for pdf
\caption[]{Variation of total multiplicity of produced $K^{-}$ with respect to $\sqrt{s_{NN}}$. Dash-dotted and dotted curve is the results obtained from HSD and UrQMD model, respectively~\cite{hsd}. Experimental data is taken from Ref.~\cite{eleven,alt1,alt2,blume1,afan,gaz}.}
\end{center}
\end{figure}

\begin{figure}
\begin{center}
\noindent
\includegraphics[height=28em]{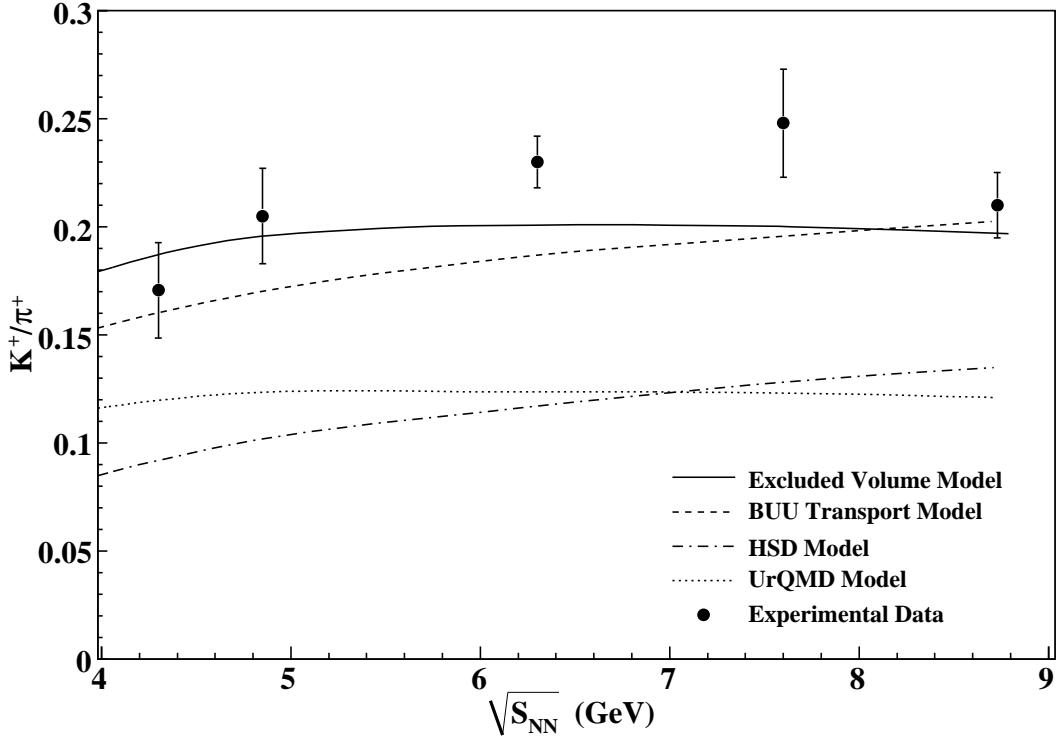}%use pdflatex for pdf
\caption[]{Variation of $K^{+}/\pi^{+}$ ratio with respect to $\sqrt{s_{NN}}$. The results obtained from HSD and UrQMD are taken from Ref.~\cite{hsd}. The results obtained from BUU transport model is extracted from Ref.~\cite{buu}. Experimental data is taken from Ref.~\cite{eleven,alt1,alt2,blume1,afan,gaz}. }
\end{center}
\end{figure}
Fig. 3 demonstrates the variation of $K^{+}$ multiplicity with respect to $\sqrt{s_{NN}}$. We have used the same freezeout volume ie., $5000~fm^{3}$ as used in the $\pi^{+}$ multiplicity. We compare our model results with the results obtained from HSD and UrQMD simulations. We have also shown the experimental data from CERN-SPS and RHIC-AGS for comparison~\cite{eleven,alt1,alt2,blume1,afan,gaz}. The results obtained from both HSD and UrQMD do not match with the data beyond the energy $5$ GeV. The overall level of agreement with the excluded volume model results is quite good. However, HSD results also suitably describe the data below  $5$ GeV. It should be emphasized here that the thermal statistical model should not be used at lower energies since number of produced particles is very small.

In Fig. 4, we presents the  variation of total multiplicity of $K^{-}$ with $\sqrt{s_{NN}}$. The results obtained from our model is in excellent agreement with the data~\cite{eleven,alt1,alt2,blume1,afan,gaz}. However, HSD and UrQMD results satisfy the experimental data only below $5$ GeV. Above $5$ GeV the total multiplicity of $K^{-}$ obtained from HSD and UrQMD are relatively small and agrees with the data. 

Fig. 5 demonstrates the variation of $K^{+}/\pi^{+}$ ratio with respect to $\sqrt{s_{NN}}$. We have compared our model results with the experimental data obtained from SPS experiment~\cite{eleven,alt1,alt2,blume1,afan,gaz}. We have further compared them with the other results like HSD, UrQMD and Boltzmann-Uehling-Uhlenbeck (BUU) transport~\cite{buu} models. We find that both the HSD as well as UrQMD model fail to give agreement with the experimental data. However, BUU model shows better results in comparison to HSD and UrQMD. Our model yields better results in agreement with the experimental data. The authors in Ref.~\cite{hsd} have suggested that the overestimation of $\pi^{+}$ multiplicity in thermal models give theoretical curve lying below the experimental data. The fireball volume used in the excluded volume model can also account for disagreement. Here we use the same volume for the production of all hadrons. CBM experiment will definitely provide an important insight in understanding the strange particle production mechanism and more vitally address the question of the existence of this ``horn'' like behaviour in the lower energy region.  

In Fig. 6, we present the variation of $K^{-}/\pi^{-}$ with respect to $\sqrt{s_{NN}}$. Our model suitably describes the data. However below $5$ GeV, again we notice disagreement. HSD and UrQMD both provide complete disagreement with the experimental data almost in the entire energy range ie., from $3$ to $9$ GeV.

\begin{figure}
\begin{center}
\noindent
\includegraphics[height=28em]{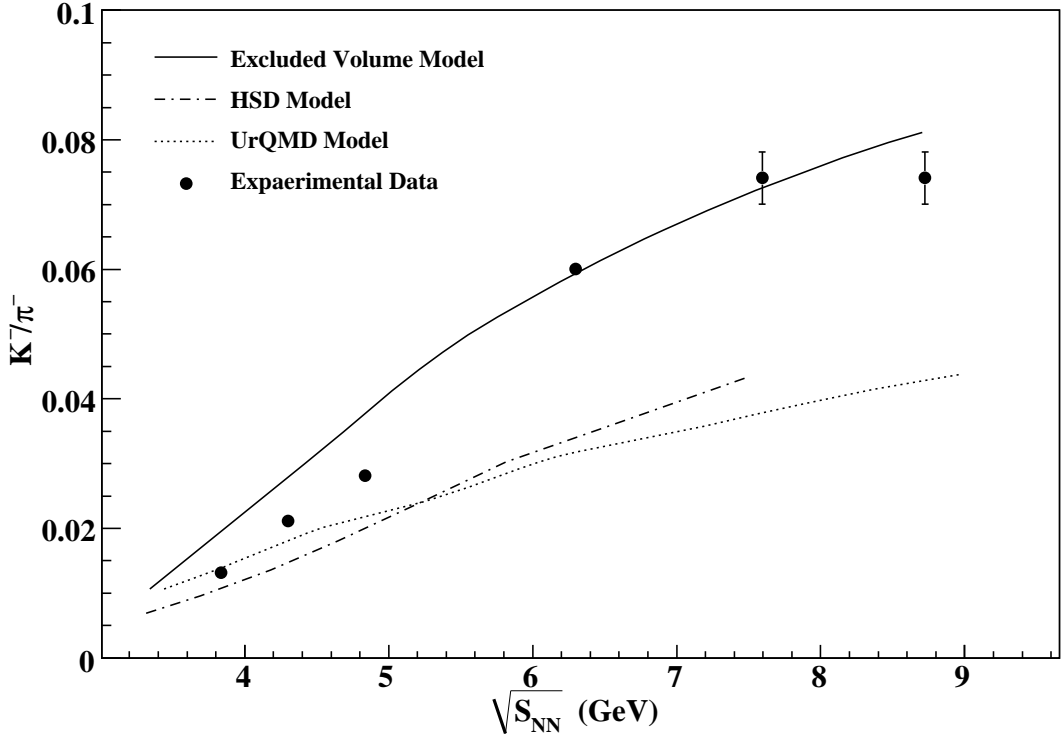}%use pdflatex for pdf
\caption[]{Variation of $K^{-}/\pi^{-}$ ratio with respect to $\sqrt{s_{NN}}$. The results obtained from HSD and UrQMD are taken from Ref.~\cite{hsd}. Experimental data is taken from Ref.~\cite{eleven,alt1,alt2,blume1,afan,gaz}. }
\end{center}
\end{figure}

%\begin{figure}
%\begin{center}
%\noindent
%\includegraphics[height=28em]{lambda_pi.eps}%use pdflatex for pdf
%\caption[]{}
%\end{center}
%\end{figure}

\begin{figure}
\begin{center}
\noindent
\includegraphics[height=28em]{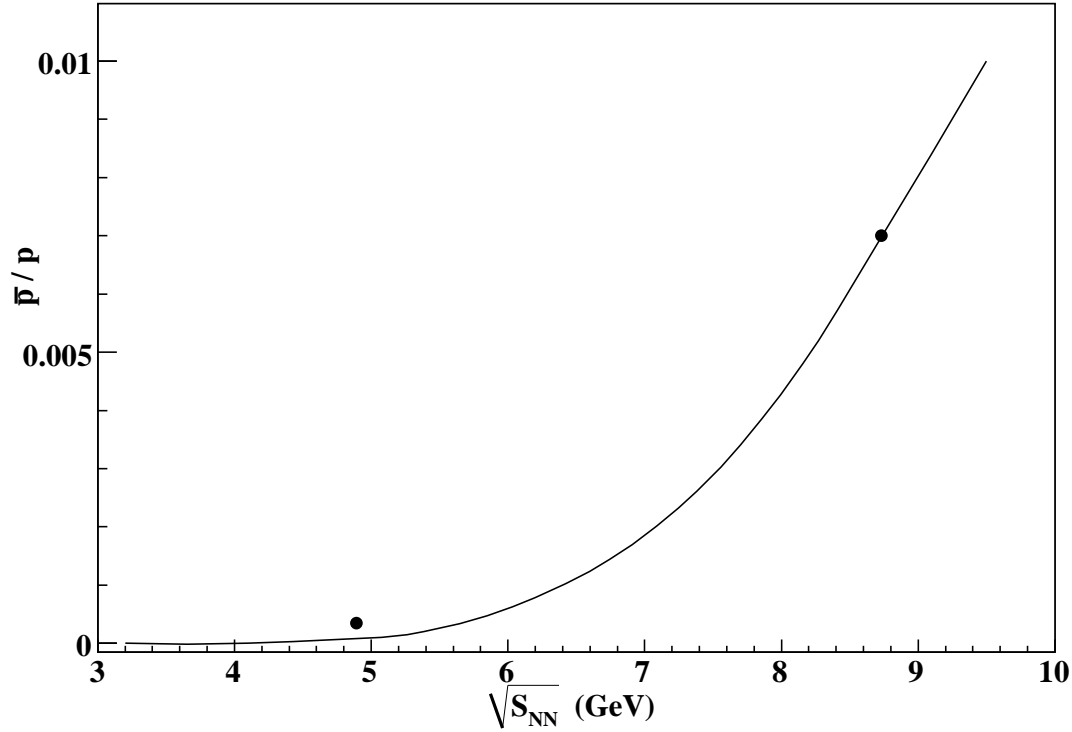}%use pdflatex for pdf
\caption[]{Variation of $\bar{p}/p$ with respect to $\sqrt{s_{NN}}$. Experimental data is taken from Ref.~\cite{eleven,alt1,alt2,blume1,afan,gaz}.}
\end{center}
\end{figure}

In heavy-ion collisions, the enhanced production of antiparticles are conjectured as indicators for the formation of deconfined QGP~\cite{G}.This would explain why the values of the ratio of antiparticle-to-particle in nucleon$\ -$nucleon (pp) collisions are higher than in heavy-ion collisions. Therefore, the initial conditions and formation time can be reflected by the surviving antiparticle. The antiparticle-to-particle ratios can be used to study particle (or antiparticle) transport and production and thus would have significant cosmological and astrophysical consequences.

 In Fig. 7 and 8, we have shown the variations of $\bar{p}/p$ and $K^{-}/K^{+}$ with $\sqrt{s_{NN}}$, respectively as obtained in excluded volume model with the center of mass energy. In both the cases, the production of antiparticle to particle is very less at lowest CBM energy. However, the asymmetry in $\bar{p}$ and $p$ production is larger as has been observed in between $K^{-}$ and $K^{+}$. As the energy increases the production of antiparticle over particle increases since there is an increase in nuclear transparency. However, it is important to state here that excluded volume model does not agree with the experimental data for thr net nucleon density at RHIC highest energy~\cite{skt}. 
\begin{figure}
\begin{center}
\noindent
\includegraphics[height=28em]{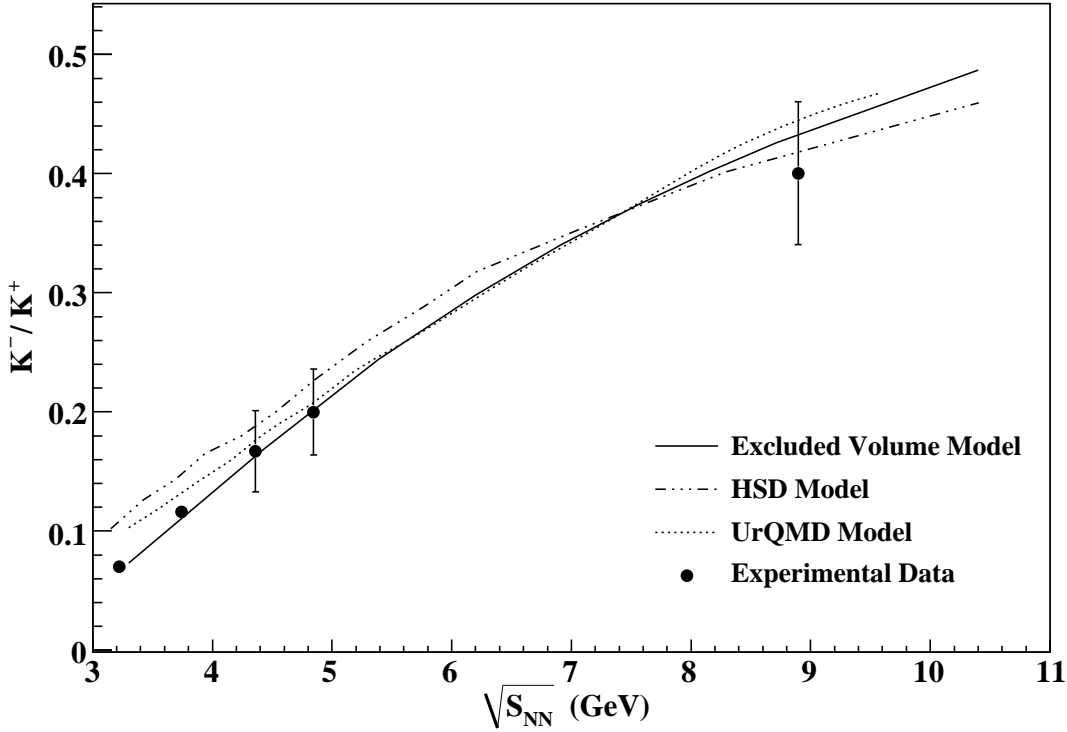}%use pdflatex for pdf
\caption[]{Variation of $K^{-}/K^{+}$ with respect to $\sqrt{s_{NN}}$. The results obtained from HSD and UrQMD are taken from Ref.~\cite{hsd}. Experimental data is taken from Ref.~\cite{eleven,alt1,alt2,blume1,afan,gaz}.}
\end{center}
\end{figure}

%\begin{figure}
%\begin{center}
%\noindent
%\includegraphics[height=28em]{kpl_pipl.eps}%use pdflatex for pdf
%\caption[]{}
%\end{center}
%\end{figure}

%\begin{figure}
%\begin{center}
%\noindent
%\includegraphics[height=28em]{lambda_pimi.eps}%use pdflatex for pdf
%\caption[]{}
%\end{center}
%\end{figure}

%\begin{figure}
%\begin{center}
%\noindent
%\includegraphics[height=28em]{phi_kmi.eps}%use pdflatex for pdf
%\caption[]{}
%\end{center}
%\end{figure}

%\begin{figure}
%\begin{center}
%\noindent
%\includegraphics[height=28em]{p_piplus.eps}%use pdflatex for pdf
%\caption[]{}
%\end{center}
%\end{figure}

In summary, we have calculated the net baryon density at freezeout in the CBM energy range which comes out to be the maximum achievable density in heavy ion collisions. Further, we have calculated the total multiplicities of various produced hadrons e.g. $\pi^{+},~K^{+},~K^{-}$ using a constant freezeout volume which is equal to $5000~fm^{3}$ same for all the hadrons. We have also calculated the ratio of $K^{+}/\pi^{+}$ and $K^{-}/\pi^{-}$ in CBM energy range. Almost all the models fail to reproduce the ``horn'' in $K^{+}/\pi^{+}$ ratio. Furthermore, we have also calculated the particle to antiparticle ratio like $\bar{p}/p$ and $K^{-}/K^{+}$. In CBM energy range these ratios increases rapidly. However, due to lack of experimental data, we do not have any guidance to precisely understand the asymmetry between hadron and anti-hadron production in heavy ion collisions.

 We conclude that our excluded volume model gives a better agreement with the available experimental data in comparison to other models. In the CBM energy range most of the multiplicities as well as particle ratios show somewhat peculiar behaviour. However, there is scarcity of experimental data. The data available suffer from poor statistics also. The upcoming CBM experiment having a high luminosity beam will provide a unique opportunity to perform systematic and comprehensive measurements, with better statistics, of bulk and rare particles. This will help us to understand the particle production mechanism and also possibly find the existence of QGP. \\
\section {Acknowledgments}
 The authors gratefully acknowledge Prof. C. P. Singh for many stimulating and inspiring discussions on the manuscript. One of the authors PKS is grateful to the University Grants Commission (UGC), New Delhi for providing a research fellowship. AP and BKS are grateful to the Department of Science and Technology (DST), Govt. of India, New Delhi for providing a research grant.

\pagebreak

\end{document}